# THE EFFECT OF VIBRATIONS ON MARANGONI CONVECTION AND MELT MIXING DURING CRYSTAL GROWING BY THE CZOCHRALSKI METHOD

A. I. Fedyushkin[*] and N. G. Bourago

Institute for Problems in Mechanics of RAS, Russia

## ABSTRACT

Vibrational melt flows in Czochralski crystal growth are investigated numerically on the basis of unsteady Navier-Stokes-Boussinesq formulation for incompressible fluid. The finite element code ASTRA is used for calculations.

It is found that the vibrations provide much more effective mixing of the melt flow compared to the rotation of the crystal and the crucible. Numerical modeling indicates the existence of standing vibrational waves on the free melt surface.

It is demonstrated that the vibrations can be used to weaken and to compensate the influence of the thermo-capillary Marangoni convection for normal and microgravity environments.

## INTRODUCTION

In Czochralski crystal growth, the thermal conditions correspond to unstable temperature distribution: cold solid-liquid interface is situated above the hot melt. In reality, there exist the intensive convective mixing of the melt due to side and bottom heating and thermo-capillary convection. This convective mixing essentially redistributes the dopant in the melt. Besides the rotation of the crystal and the crucible effects the mixing and many studies are devoted to the rotating flows in Czochralski crystal growth, but only recently the investigation of the vibrational flows started (see, for instance [1]). For more deep understanding the influence of vibrations on hydrodynamics, heat and mass transfer in Czochralski crystal growth it is necessary to study not only vibrational flow itself, but also its interaction with other mentioned above types of the flows. The scientific group of Prof. E.V. Zharikov investigates the influence of the vibrations on crystal growth experimentally [2].

---

*Corresponding author, Alexey Fedyushkin fai@ipmnet.ru

In papers [3, 4] the vibration flows mores studied numerically for vertical Bridgman crystal growth. The goal of the present paper is to numerically investigate the convective heat and mass transfer in Czochralski crystal growth taking into account the thermal, thermo-capillary and forced convection in presence of the vibrating crystal.

FORMULATION OF PROBLEM

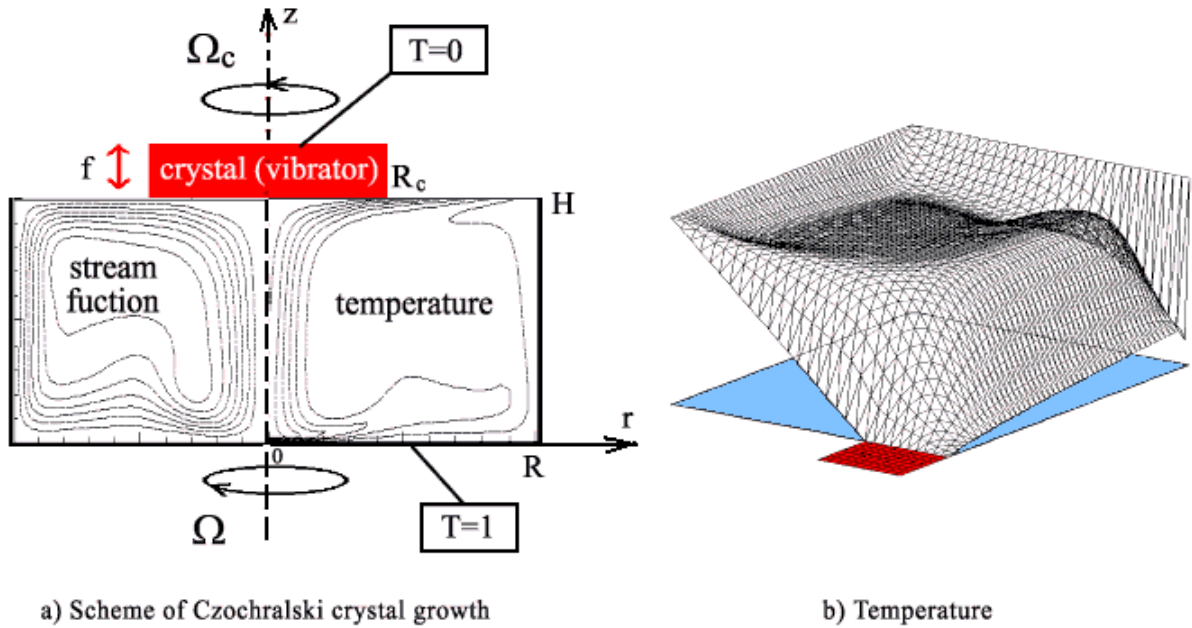

a) Scheme of Czochralski crystal growth                    b) Temperature

Fig.1 a) Computational region; Natural convection:
on left side - stream function, on right side - themperature; b)Temperature
(Gr = 214610, Pr = 5.43, H/R = 1, $R_c$/R = 0.3)

The computational region is shown in Fig. 1, where growing crystal plays the role of the vibrator too, R - a radius of the crucible, $R_C$ - a radius of the crystal, H - a height of the crucible, 0z - an axis of symmetry.

The following assumptions have been in use: axial symmetry of Czochralski crystal growth and are permanency of growth rate and thermal boundary conditions. The vibrations have a small amplitude so the displacements of vibrating crystal are negligible and only the vertical velocity of vibrating solid-liquid interface is predefined as a harmonic function $v = A\omega \sin(\omega t)$, where A and $\omega$ - an amplitude and a frequency, correspondingly.

The used from of Navier-Stokes-Boussinesq equations read:

$$\frac{\partial u}{\partial r} + \alpha \frac{u}{r} + \frac{\partial w}{\partial z} = 0$$

$$\frac{du}{dt} - \alpha \frac{v^2}{r} = -\frac{1}{\rho_0} \frac{\partial p}{\partial r} + \frac{1}{r^\alpha} \frac{\partial}{\partial r}\left(r^\alpha \nu \frac{\partial u}{\partial r}\right) + \frac{\partial}{\partial z}\left(\nu \frac{\partial u}{\partial z}\right) - \alpha \nu \frac{u}{r^2}$$

$$\frac{dw}{dt} = -\frac{1}{\rho_0} \frac{\partial p}{\partial z} + \frac{1}{r^\alpha} \frac{\partial}{\partial r}\left(r^\alpha \nu \frac{\partial w}{\partial r}\right) + \frac{\partial}{\partial z}\left(\nu \frac{\partial w}{\partial z}\right) + g\beta_T(T - T_0)$$

$$\frac{dv}{dt} + \alpha \frac{uv}{r} = \alpha \left[\frac{1}{r}\frac{\partial}{\partial r}\left(r\nu\frac{\partial u}{\partial r}\right) + \frac{\partial}{\partial z}\left(\nu\frac{\partial u}{\partial z}\right) - \nu\frac{u}{r^2}\right]$$

$$\frac{d\rho c_p T}{dt} = \frac{1}{r^\alpha}\frac{\partial}{\partial r}\left(r^\alpha \lambda \frac{\partial T}{\partial r}\right) + \frac{\partial}{\partial z}\left(\lambda \frac{\partial T}{\partial z}\right)$$

Where: u and w are velocities in r and z directions, v is an azimutal velocity, T is a temperature, p is a pressure, g is gravity acceleration, $\beta_T$ is thermal expansion factor, $\nu$ is kinematic viscosity factor, $\rho_0$ - is a density, $c_p$ - is heat capacity at constant pressure, $\alpha$ is a geometry factor, which equals to 0 for flat geometry or to 1 for axisymmetrycal geometry, $\lambda$ a heat conductivity.

The boundary conditions read:

at the axis of symmetry

$$r = 0 \;,\; u = 0,\; \frac{\partial w}{\partial r} = 0,\; v = 0,\; \frac{\partial T}{\partial r} = 0\,;$$

at the surface of the crystal

$$z = 0 \;,\; u = 0,\; \mathbf{w = W_s + A\,\omega\,\sin(\omega\,t)},\; v = 2\pi r\Omega_C,\; T = 0\,;$$

On the wall of the crucible

$$r = R\,,\; u = 0,\; w = 0,\; v = 2\pi R\Omega,\; T = 1\,;$$

at the free surface of the melt

$$\mu\frac{\partial u}{\partial z} = -\frac{\partial \sigma}{\partial r},\; \frac{\partial w}{\partial z} = 0,\; \frac{\partial v}{\partial z} = 0,\; \frac{\partial T}{\partial n} = 0 \text{ or } T = (r - R_C)/(R - R_C)\,,$$

where $W_s$ - crystal growth rate, A - an amplitude, $\omega$ - a frequency of vibrations, $\Omega_C$ - a frequency of rotation of crystal, $\Omega$ - a frequency of rotation of crucible, R - radius of crucible, $R_C$ - radius of crystal, $\sigma$ is a surface tension factor, $\beta_\sigma = \frac{\partial \sigma}{\partial T}$, $\mu$ is a viscosity factor.

Formulation of boundary conditions has principal meaning for correctness of the model and comparison with experimental data. In general case boundary conditions can vary in time.

Initial conditions read: $t = 0$, $u = 0$, $w = 0$, $v = 0$, $T = 0$.

The problem characterized by following parameters of similarity: Prandtl number $Pr = \nu c_p / \lambda$, Reynolds number $Re_\Omega = \Omega R^2 / \nu$, $Re = W_s R / \nu$, Grashof number $Gr = g\beta\Delta T R^3 / \nu^2$, (or Rayleigh number Ra=GrPr) and Marangoni number $Mn = \sigma\beta_\sigma R\Delta T / \mu\nu$. In most cases the parameters take the values: $Pr = 5.43$, $Re_\Omega$ and $Re < 10^3$, $Gr = 0 - 10^6$, $|Mn| = 0 - 500$. The harmonic vibrations of the crystal have the amplitude A=100μm and frequencies $f = \omega / 2\pi = 0$-100 Hz. The rate of crystal growth for all runs is the same and equal to $W_s = 0$ or 0.3 cm/h.

## NUMERICAL METHOD

The most essential features of used numerical method can be described in a following way. For a typical convection-diffusion equation

$$\frac{\partial \mathbf{A}}{\partial t} + \mathbf{u} \cdot \nabla \mathbf{A} = \nabla \cdot (k\nabla \mathbf{A}) + \mathbf{F}$$

The following variational implicit Bubnov-Galerkin scheme is used:

$$\int_V \left( \frac{A^{n+1} - A^n}{\Delta t^n} + \mathbf{u}^n \cdot \nabla A^{n+1} \right) \left( \delta A + \Delta t^n \mathbf{u}^n \cdot \nabla \delta A \right) d\mathbf{V} +$$

$$+ \int_V k_1 \nabla A^{n+1} \cdot \nabla \delta A dV = \int_V F^{n+1} \delta A dV + \int_S k\mathbf{n} \cdot \nabla A^{n+1} \delta A d\mathbf{S}$$

Here an exponential viscosity correction a la Samarski is used to provide monotonous behavior of solution:

$$k_1 = k \left( 1 - \frac{0.5 \max(Uh, U^2 \Delta t)}{k} \right)^{-1}$$

Linear finite elements in space were used. Auxiliary algebraic problems were solved by non-matrix conjugate gradients method with preconditioning by using diagonal approximation of stiffness matrix. Algorithm is unconditionally stable but for good accuracy time step should not differ much from the value of Courant's time step: $0.1\Delta t_c < \Delta t^n < 10\Delta t_c$, where $\Delta t_c$ - is a convectional (Courand's) time step. Incompressibility was handled by penalty method (first method), by pressure correction (second method), by the vorticity-stream function formulation (third method) and by Chorin's artificial compressibility (forth method). Results are in a good accordance for all four implements techniques. The algorithms included incorporated into known hydrocode "ASTRA" for 2D and 3D geometry [5-7].

The characteristics of the AVF (Averaged Vibrational Flow) were defined by using the following formula:

$$F_{average} = \frac{1}{t} \int_0^t F \, dt \, ,$$

where t - time, F - some of the flow characteristics

The following two issues should be underlined: firstly, the value of time step was chosen to have at least 20 per period, secondly, the presented results correspond to the time, when the AVF becomes quasi-stationary, if the opposite is not stated.

RESULTS OF CALCULATIONS

The Fig. 1a shows the scheme of the solution domain, the stream function isolines (left hand side picture) and the isotherms (on the right) for the case of thermal convection are shown there also (Gr=2 $10^5$, Pr=5.43, H/R=1, $R_c$/R=0.3). In Fig. 1b the spatial graphs of the temperature field can be seen. The convection forms the thermal boundary layers and levels the temperature in the kernel of the domain (Fig. 1b). However, its action is so strong as the action of the vibrations what is demonstrated farther.

On the free melt surface, the Marangoni convection effect can take place. Its intensity depends on the temperature gradient near the free surface. Fig. 2 shows the results for Marangoni convective flow (Mn=-500) under two types of temperature boundary condition on the free surface. The left column corresponds to the case of temperature linear variation from 0 to 1, the right column responds to the case of thermal isolation. In the second case, the temperature profile and the temperature gradient on the free surface are formed due to the melt motion and the heat conductivity.

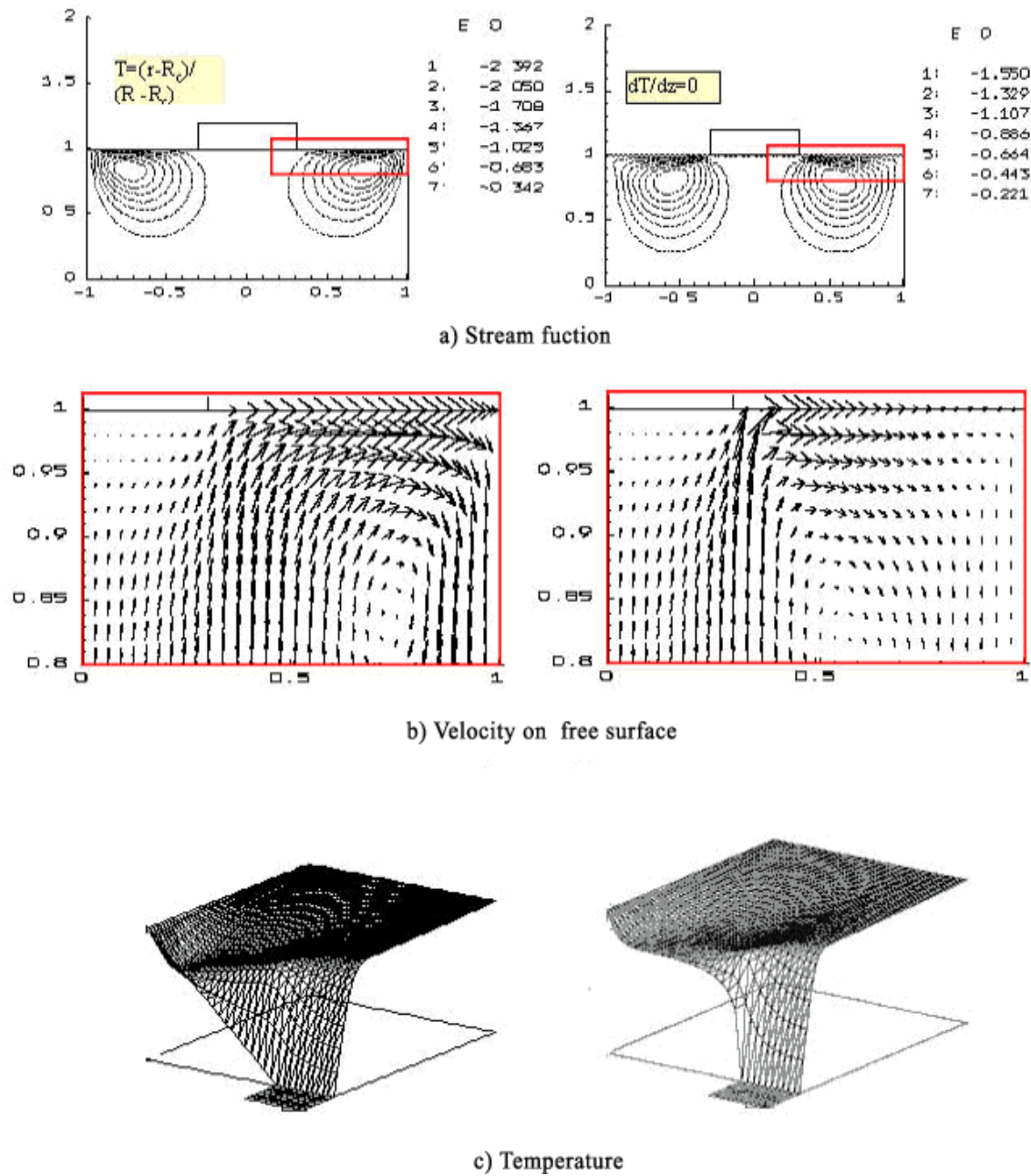

a) Stream fuction

b) Velocity on free surface

c) Temperature

Fig.2. Marangoni convection (Mn=-500)

Therefore, the flows, produced by the Marangoni convection and the temperature distribution in these two cases are different. Fig. 2a demonstrates the isolines of the stream function. It is seen that the maximal values of the stream function and the velocities on the free surface in these two cases are situated in opposite sides respectively the middle point of the free surface. Fig. 2b shows the velocity field in the narrow zone near the free surface (in Fig. 2a this

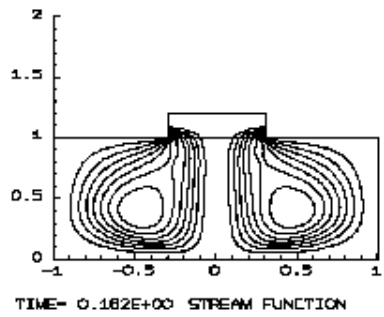

a) Stream function

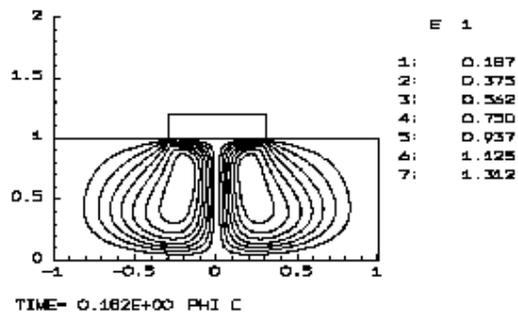

b) Stream function of AVF

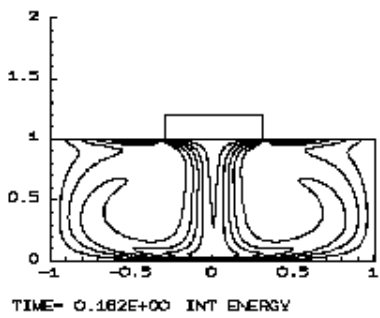

c) Temperature

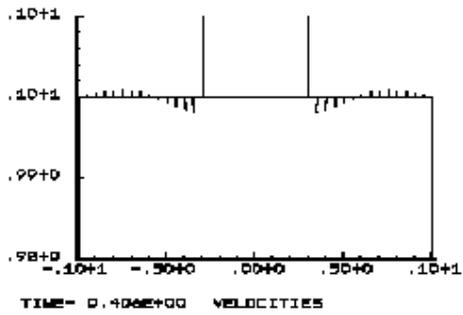

d) Average velocity on the free surface

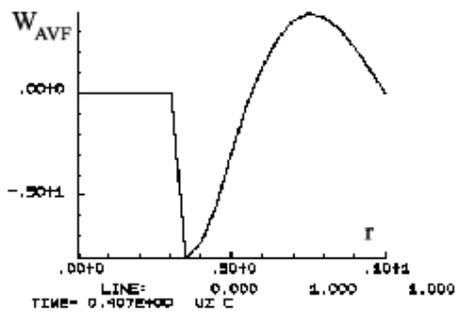

e) Profile of the average velocity (z=1)

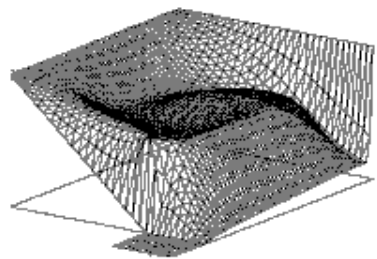

f) Temperature

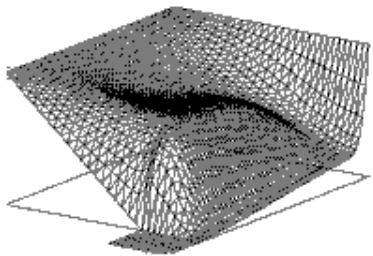

g) Temperature

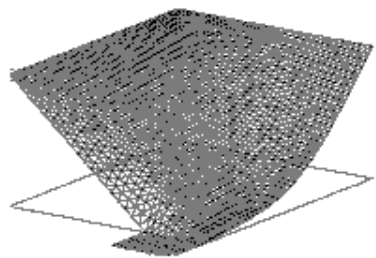

h) Temperature

Fig. 3. Vibration of the crystal a)-f) f=50 Hz, g) f=10 Hz, h) f=1 Hz

zone is marked by frame). From the distribution of the velocities on the free surface it is seen that the Marangoni convection has lower intensity in the case of thermal isolation. The temperature distribution in the domain and the temperature profiles on the free surface are presented in Fig. 2c.

In the first case (linear temperature profile on the free surface) the constant horizontal temperature gradient is predefined and therefore the velocity of the Marangoni convection is increasing on the part from the crystal to the side wall. In the second case, the temperature on the free surface is leveled because of the Marangoni convection and becomes constant over almost the entire free surface. The zones with high level of gradients are situated near the crystal and near the side wall of the crucible. So even in the simplest case the regimes and the character of the convective flows may be different. In addition, it should be pointed out that similar mechanism of the boundary condition influence acts in the case of thermal convection as well.

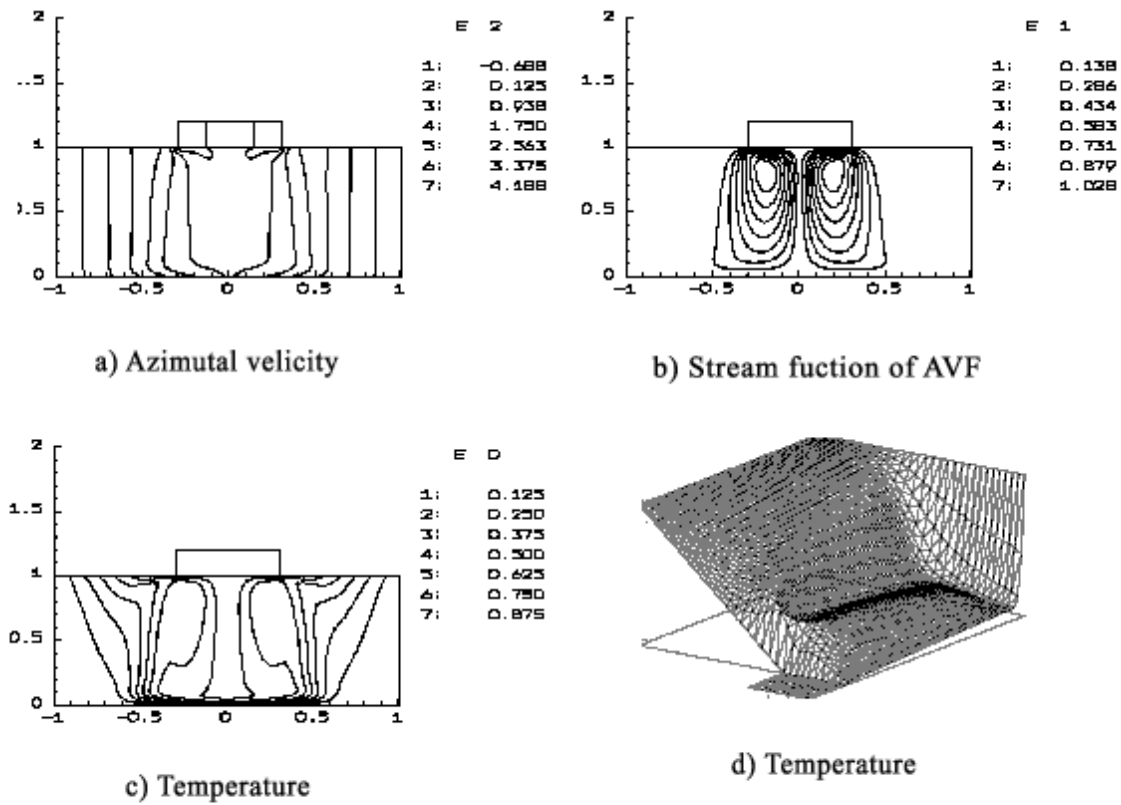

a) Azimutal velicity

b) Stream fuction of AVF

c) Temperature

d) Temperature

Fig. 4. Vibration and rotation $(f=50Hz, Re_{crystal} = -500, Re_{crucible} = 500)$

The Marangoni convection increases the vertical temperature gradient at the solid-liquid interface compared to the case of pure thermal conductivity.

The Fig. 3 shows what kind of flows are forced by the vibrations of the crystal (f=50, 10, 1 Hz) and what kinds of the temperature distributions appear. The Fig.3 shows: the isolines of instant stream function (Fig.3a), isolines of the AVT stream function (Fig. 3b), isotherms (Fig. 3c), the AVT velocity on the free surface (Fig. 3d), the profile of the AVT vertical velocity on the free surface (Fig. 3e). The acsonometric projections of the temperature field for the three values of the frequency (f=50, 10, 1Hz) are depicted in Fig. 3(f, g, h). As it is seen the vibrations lead to intensive melt mixing and to narrowing the temperature boundary layers at the bottom of the crucible and near the crystal.

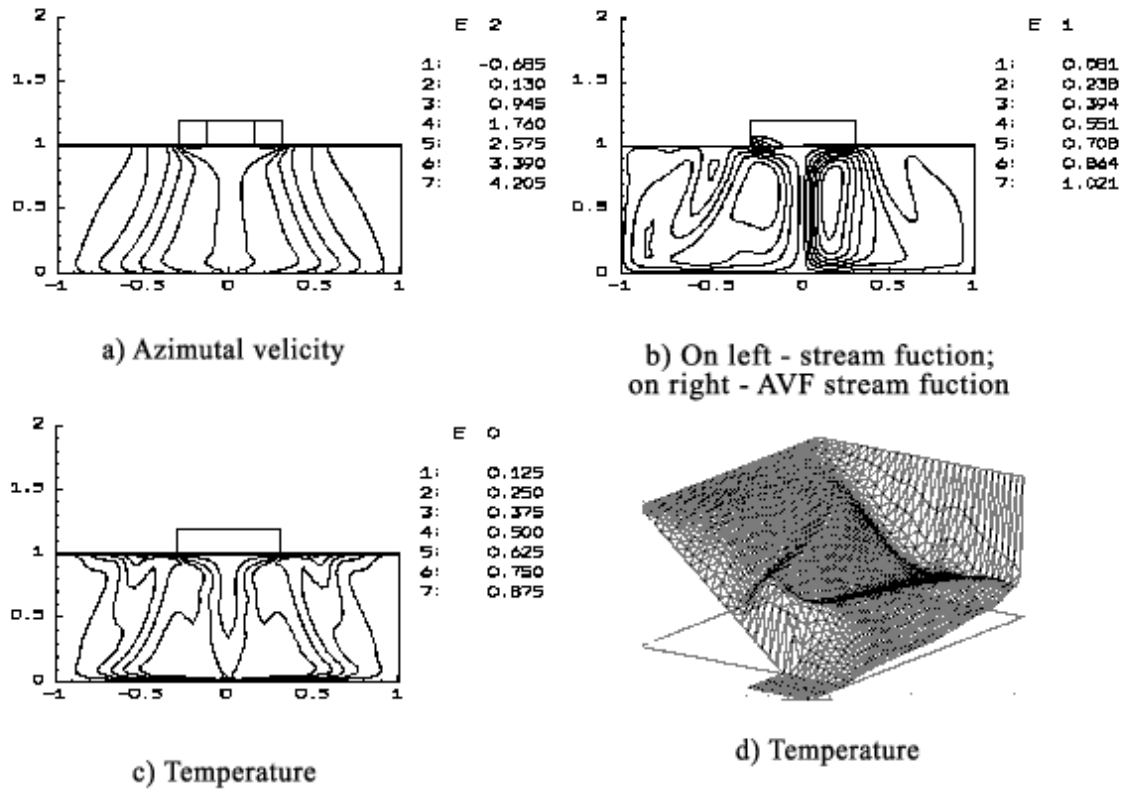

Fig. 5. Vibration, rotation and natural convection
(A= 100 μm, f = 50 Hz, $Re_{crystal}$ = -500 , $Re_{crucible}$ = 500, Gr = 214610, Pr = 5.43)

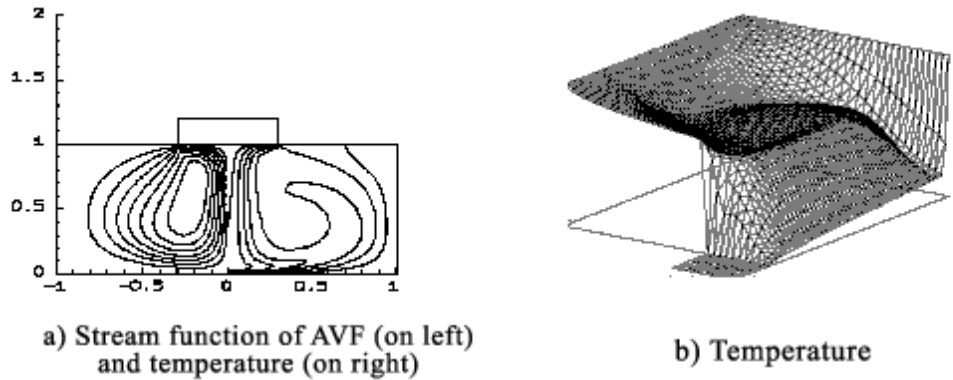

a) Stream function of AVF (on left)
and temperature (on right)

b) Temperature

Fig. 6. Vibration and Marangoni convection

(f=50 Hz, Mn=-500, insulation condition at free surface)

The isolines of the AVF stream function are similar to the isolines of stream function for the case of crucible rotation. The temperature field is varied in the same way as in the case of thermal convection. Nevertheless, already under frequency f=50hz the vibrations level the temperature field inside the crucible.

The vibrations cool the melt flow kernel stronger than the thermal convection. Variation of the temperature is observed only near the walls of the crucible. On the free surface the distribution of averaged velocity shows the presence of standing waves. They are observed in the experiments [2] also.

Comparing the temperature fields in Fig. 1b and Fig. 3f one can conclude that the vibrations of frequency f=50 Hz cause better melt mixing and temperature leveling in the center of the flow domain compared to the thermal convection (Gr=2 $10^5$, Pr=5.43). It means that this effect is independed on gravity. The vibrations can be a good mixing mechanism in crystal growth production technologies and in Space, and on the Earth.

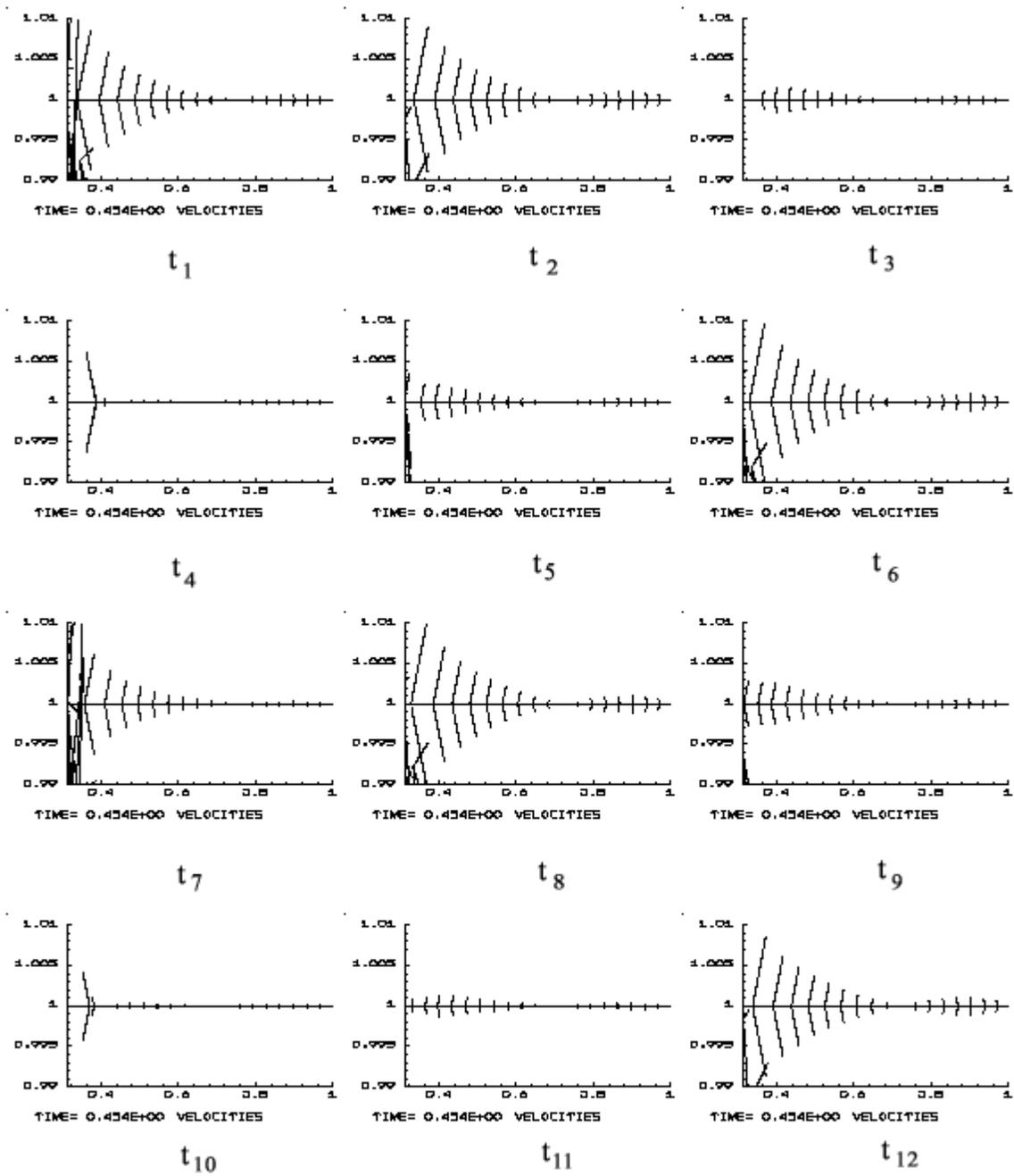

Fig. 7. Velocity on the free surfer for the different moments of the time
(A=100μm, f=50Hz, Mn=500, insulated free surface)

In the Czochralski crystal growth besides the convection mixing mechanism the rotation of the crystal and the crucible also can be used. The crystal and the crucible are often rotating in opposite directions. The rotation of crucible as well as the thermal convection cools the domain under the crystal, best it happens in a different manner. The rotation of the crystal leads to the melt flow in clockwise direction. In the case of rotating crystal the domain under the crystal is heated. The width of thermal boundary layer near the solid-liquid interface becomes smaller with acceleration of rotation (with increase of rotational Reynolds number). Fig. 4 indicates the result of the joint action of vibrations (f=50 Hz) and opposite rotations of the crystal and the crucible ($Re_{crystal}$=-500, $Re_{crucible}$=500). Fig. 4 shows the isolines of the azimutal velocity (a), the isolines of the AVF stream function (b), the isotherms (c) and the temperature field (d).

The vibrations of crystal (f=50 Hz) and rotation essentially vary the temperature under the crystal. The width of the temperature boundary layer becomes smaller compared to the case of absence of rotation. Cool zone is situated under the crystal as a column connecting the crystal and the bottom of crucible, on the right of this zone the homogeneous hot zone is appeared. Near the crystal and near the bottom of crucible the strong boundary layers exist.

Fig. 5 shows that the action of the vibration, the rotation and the thermal convection leads to more complex structure of the flow. Conditionally it can be divided into two domains by the line, which connects the external side of the crystal and the bottom corner of the crucible. The flow structure in the external domain is formed by the thermal convection. In the internal domain the flow structure is formed by the vibration and by the rotation. The fight between the vibration and the rotation is observed. This fight can lead to the homogeneity alterations of the temperature field.

Fig.6 shows the interaction between the vibrations and the Marangoni convection for the case of the thermo-isolated free surface. The vibrations make the temperature distribution at the free surface more homogeneous. This decreases the temperature gradient along the free surface (Fig. 6b) and makes lower the Marangoni convection intensity.

Fig. 7 shows the influence of the vibrations on the velocities near the free surface for various instants during one period. It can be seen that on the free surface the zones with zero, negative and positive values of the tangential velocity exist. This confirms the effect of standing waves existence, which are observed in the experiments.

The results obtained in this work are consistent and addendum the results on the influence of vibrations, both for the Czochralski method [8] and for the Bridgman technique [9, 10].

## CONCLUSIONS

It is shown that the vibrations can decrease the width of boundary layers and increase the temperature gradients. This can intensify heat and mass transfer near the solid-liquid interface and increase the rate of crystal growth.

It is shown that the vibrations can be used for effective mixing of the melt.

The existence of standing waves on free melt surface under influence of vibrations is numerically confirmed.

It is shown also that the vibrations can be used to weaken the influence on the thermo-capillary Marangoni convection.

## REFERENCIES

1. G.Z.Gershuni and D.V.Lubimov, (1998) Thermal vibrational convection. John Willey&Sons Ltd, 357p.
2. E.V.Zharikov, L.V.Prihod'ko, N.R.Storozhev, (1990) Fluid flow formation resulting from forced vibration of a growing crystal. J.Crystal Growth, 99, pp.910-914.
3. A.I. Fedyushkin, N.G. Bourago, (2001) Influence of vibrations on boundary layers in Bridgman crystal growth. Proceedings of $2^{nd}$ Pan Pacific Basin Workshop on Microgravity Sciences 2001, paper CG-1073.
4. A. I. Fedyushkin, N. G. Bourago, V. I. Polezhaev and E.V. Zharikov, (2001) Influence of vibration on heat and mass transfer during crystal growth in ground-based and microgravity environments. Proceedings of $2^{nd}$ Pan Pacific Basin Workshop on Microgravity Sciences 2001, paper CG-1065.
5. N. G. Bourago and V. N. Kukudzhanov, (1988) Numerical Simulation of Elastic Plastic Media by the Finite Element Method, Pre-print IPMech AS USSR, N.326, pp. 1-63.Second edition in "Computer Mechanics of Solids", issue 2, 1991,pp. 78-122.
6. N.G.Bourago, (1994) Numerical methods for non-linear processes in elastic plastic media. In "Lechers of FEM-94 Seminar", Chalmers Univ. Techn. Struct.Mech.Dept., Publ: 1994.1, Gothenberg, pp. 1-15.
7. Burago N. G., Fedyushkin A. I. Numerical solution of the Stefan problem // Journal of Physics: Conference Series. 2021. V. 1809, No. 1. 012002. DOI: http://dx.doi.org/10.1088/1742-6596/1809/1/012002.
8. Fedyushkin A.I. Heat and mass transfer during crystal growing by the Czochralski method with a submerged vibrator Journal of Physics: Conference Series, Institute of Physics (United Kingdom), V. 1359, 012054, 2019. DOI: 10.1088/1742-6596/1359/1/012054
9. Fedyushkin A.I., Burago N.G., Puntus A.A. Convective heat and mass transfer modeling under crystal growth by vertical Bridgman method. Journal of Physics: Conference

Series, IOP Publishing ([Bristol, UK], England), V. 1479, 012029, 2020. DOI: 10.1088/1742-6596/1479/1/012029.

10. Fedyushkin A.I., Burago N.G., Puntus A.A. Effect of rotation on impurity distribution in crystal growth by Bridgman method. Journal of Physics: Conference Series, Institute of Physics (United Kingdom), V. 1359, 012045, 2019. DOI: 10.1088/1742-6596/1359/1/012045